\documentclass[10,prl,aps,superscriptaddress,twocolumn]{revtex4-2} 
\UseRawInputEncoding
\usepackage[T1]{fontenc}
\usepackage[utf8]{inputenc}
\usepackage{hyperref}
\usepackage{graphicx}    
\usepackage{caption}     
\usepackage{subcaption}  
\usepackage{color}
\usepackage{bm}
\usepackage{epstopdf}
\usepackage{amsfonts}
\usepackage{amsmath}
\usepackage{bbold}
\usepackage{cancel}
\usepackage{natbib}
\usepackage{amssymb}

\newcommand{\respch}[1]{\textcolor{black}{#1}}

\usepackage[textsize=scriptsize, colorinlistoftodos]{todonotes}

\newcommand{\Rubidio}{\text{Rb}}
\newcommand{\Potasio}{\text{K}}

\usepackage{tikz}

\newcommand{\tikzcircle}[2][red,fill=red]{\tikz[baseline=-0.6ex]\draw[#1,radius=#2] (0,0) circle ;}

\begin{document}

\preprint{APS/123-QED}

\title{
Vorticity Induced by Non-frontal Collisions of Quantum Droplets}
\author{J. E. Alba-Arroyo}
\email{jose.arroyo@pw.edu.pl}
\affiliation{%
Faculty of Physics, Warsaw University of Technology, Ulica Koszykowa 75, 00-662 Warsaw, Poland}%
\affiliation{%
Departamento de F\'{\i}sica Cu\'antica y Fot\'onica, Instituto de F\'{\i}sica, Universidad Nacional Aut\'onoma de 
M\'exico\\ Cd. de M\'exico C.P. 04510,
M\'exico
}%

\author{Santiago F. Caballero-Benitez}%
\email{scaballero@fisica.unam.mx}
 \affiliation{%
Departamento de F\'{\i}sica Cu\'antica y Fot\'onica, LSCSC-LANMAC, Instituto de F\'{\i}sica, Universidad Nacional Aut\'onoma de 
M\'exico\\ Cd. de M\'exico C.P. 04510,
M\'exico
}%
\author{Rocio J\'auregui}%
 \email{rocio@fisica.unam.mx}
\affiliation{%
Departamento de F\'{\i}sica Cu\'antica y Fot\'onica, Instituto de F\'{\i}sica, Universidad Nacional Aut\'onoma de 
M\'exico\\ Cd. de M\'exico C.P. 04510,
M\'exico
}%

\date{\today}

\begin{abstract} 
The rotational dynamics induced by the non-frontal binary collisions of quantum droplets composed of ultracold alkali atoms are analyzed. A theoretical study is presented within the extended Gross-Pitaevskii equation framework, using experimentally feasible conditions. Numerical experiments elucidate a rich landscape of possible topological excitations in the system that are robust towards measurements. The collision of heteronuclear quantum droplets composed of $^{41}$K and $^{87}$Rb atoms in the incompressible regime gives rise to dynamical instabilities that spontaneously generate topological defects: vortex rings, dislocation lines, and vortices in one species. Their presence depends on the Weber number and the impact parameter. An experimental proposal for vortex detection in both real and Fourier space using interaction ramps is described. 

\end{abstract}
\maketitle

A comprehensive analysis of the rotational dynamics of superfluids has enabled a better understanding of the interplay between coherent, collective, and non-equilibrium phenomena in quantum systems. As a consequence, studies on superfluidity, the evolution of quantized vortices, vortex lattices, sound waves, and persistent currents as a function of the statistics of ultracold atomic systems and their interparticle interactions,  have shown an always increasing scientific interest~\cite{RevModPhys.29.205,VortexBook,LeggettBook}.

Most experiments induce rotation in quantum-degenerate fluids via appropriate boundary conditions or external potentials. For instance, in liquid helium, the fluid can be set in motion by the surface roughness of the container walls when the container rotates rapidly enough \cite{donnelly91}. Similar techniques have been used in experiments involving ultracold alkali atomic gases with static magnetic traps and stirring lasers \cite{madison2000}. In both cases, for large enough rotation frequency, the superfluid presents lines of singularity in its phase field, that is,  vortex filaments are generated.  Here, we explore the behavior of a self-confined superfluid where the rotational dynamics are induced by its collision with another sample of such a superfluid. This scenario involves a multi-component mixture of alkali atoms in the quantum droplet regime.

\begin{figure}[t!]
    \centering
            \includegraphics[width=1.0\linewidth]{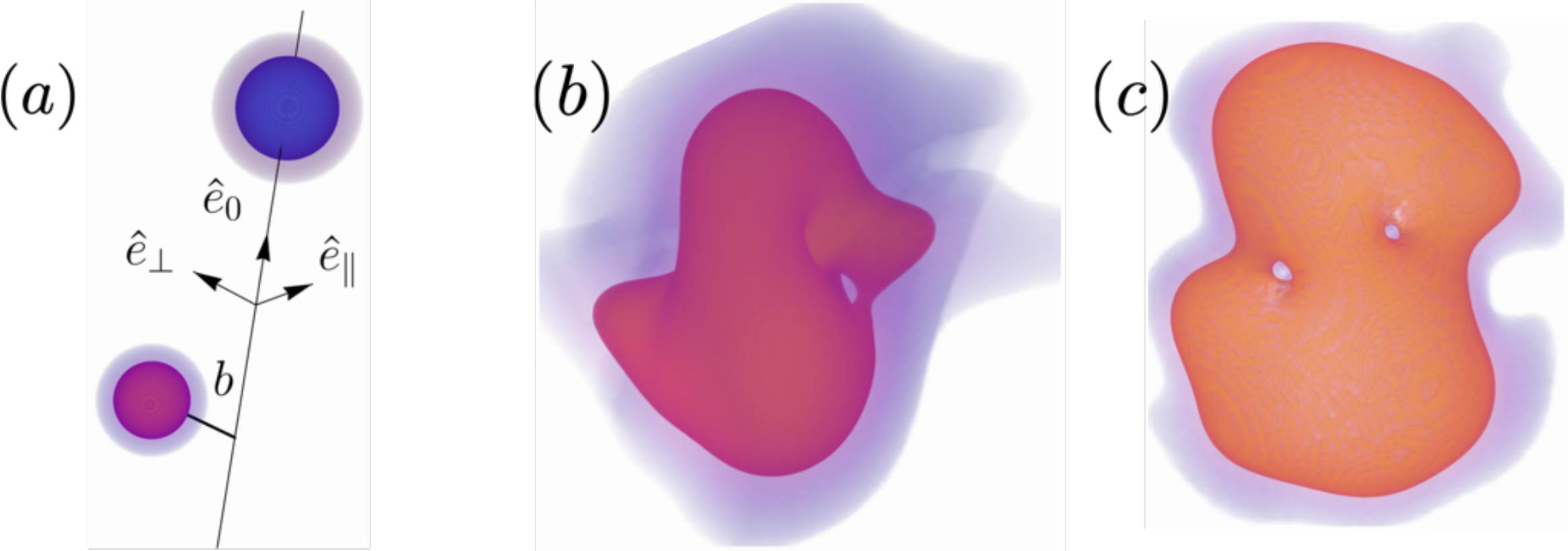} 
     \caption{
         {\bf  Diagramatic proposal for non-frontal collisions of quantum droplets and illustrative results when vortex filaments are generated}. (a) Setup for the two droplets before the collision, $\hat{e}_{0}$ defines the direction of the relative wavenumber $2k_0$ between the droplets, $\hat{e}_{\perp}$ the direction along which the impact parameter $b$ is evaluated, and, $\hat{e}_{\parallel}$ completes the local reference frame system. (b) and (c) illustrate the outcomes after the collision of the quantum droplets: (b) one vortex and (c) twin vortices in the heavy component of the degenerate gas.
         The colliding droplets are formed by $N_1=N^{\Rubidio{}}_1 + N^{\Potasio}_1=1.81\times 10^5$ and $N_2=N^{\Rubidio}_2 + N^{\Potasio}_2=1.31\times 10^5$ atoms; $N^{\Potasio}_{i=1,2}/N^{\Rubidio}_{i=1,2}=\sqrt{g_{\Rubidio \Rubidio}/g_{\Potasio \Potasio}}=0.8735$ For (b):  $\mathcal{W}e_1 \simeq \mathcal{W}e_2= 1.35$, $b/(R_{01}+R_{02})=0.33$, $R_{01}=3.28 \xi$, $R_{02}=2.81 \xi$, for (c): $\mathcal{W}e_1 \simeq \mathcal{W}e_2 =2.11$, $b/(R_{01}+R_{02})=0.33$, $R_{01}=3.28 \xi$, $R_{02}=2.81 \xi$. The evolution times are (b) $t_{\mathrm{evo}}=14.74 \tau$ and (c) $t_{\mathrm{evo}}=12.06 \tau$.
         }\label{fig:Fig1}
         \end{figure}

Mixed atomic condensates exhibit diverse interaction regimes and physics beyond the mean-field description~\cite{Petrov2015}. A good example of such a system is quantum droplets, which are self-confined due to beyond-mean-field effects preventing collapse. Current droplet experiments with ultracold atoms include dipolar Bose-Einstein Condensates (BECs) using $^{164}$Dy and $^{166}$Er atoms \cite{schmitt2016self}, homonuclear BECs mixtures using hyperfine states of $^{39}$K~\cite{Science2018Cesar,Semeghini2018} and heteronuclear BECs of $^{41}$K - $^{87}$Rb~\cite{Derrico2019} and of $^{23}$Na -$^{87}$Rb \cite{guo2021}. The relative number of each atomic species in a mixture and the coupling strength between them are crucial parameters for observing self-trapping, exhibiting self-evaporation, achieving an incompressible regime, and admitting low-energy excitations~ \cite{Science2018Cesar,Semeghini2018,PhysRevResearch.3.043139,Alba2021}.

 Vortices in both homonuclear  and heteronuclear   mixtures of superfluid atomic gases are usually  generated and described in a static framework:  the system is externally imprinted with a given amount of angular momentum, and the corresponding rotating frame of reference is used. The resulting distribution of angular momenta for non-spherical droplets is shape-dependent, favoring the nucleation of vortices in the heaviest component~\cite{VorticesAncilotto}. In $^4$He quantum vortex lattices with large cores within single nanodroplets have been observed experimentally~\cite{gomez2014shapes}; this remains elusive in alkali BECs.  Recently, it has been shown that imbalanced populations or spin-mode excitations are energetically feasible, but with a limited lifetime~\cite{PARKERIMBALANCED2025}. 

\begin{figure}[t!]
     \includegraphics[width=1.0\linewidth]{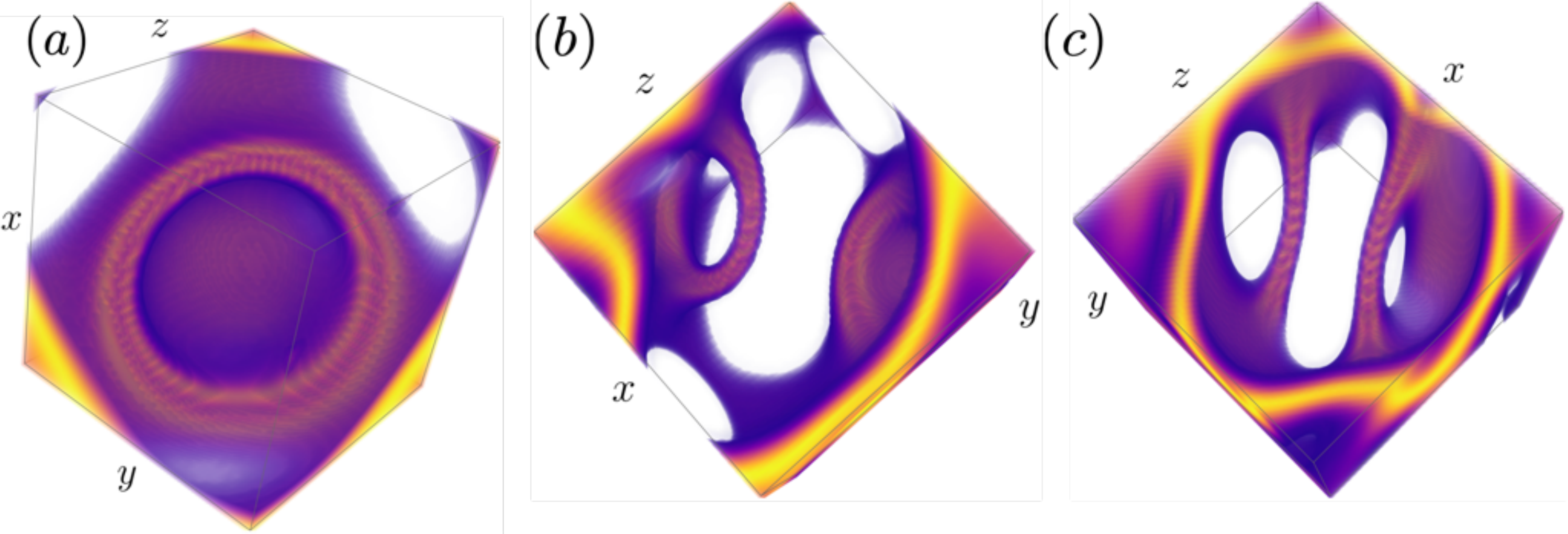}
    \caption{
    {\bf{Topological excitations generated dynamically within the atomic cloud as a result of a binary collision.}} (a): A vortex ring that has a twin ring (not shown) spatially at 180$^\circ$ degrees is formed during a collision of identical quantum droplets. The twin vortex rings are the seed  of other higher-energy excitations.  (b): One vortex with a twisted vortex ring. The kinetic energy is high enough to generate a density depression that traverses the cloud, as shown in Fig. \ref{fig:Fig1}(b). (c): Two vortices that traverse the cloud, typically generated with high kinetic energy, as shown in Fig. \ref{fig:Fig1}(c). Parameters are: (a) $N_1=N_2=1.81\times 10^5$, $\mathcal{W}e=1.19$, $b/2R_{01}= 0.154$, $R_{01}=3.18 \xi$, $\tau_{\mathrm{evo}}=10.34\tau$; for (b) and (c), the same as Fig.\ref{fig:Fig1}b and Fig.\ref{fig:Fig1}c.}
    \label{fig:Fig2}
\end{figure}

\begin{figure}[t!]
    \centering   
     \includegraphics[width=1.0\linewidth]{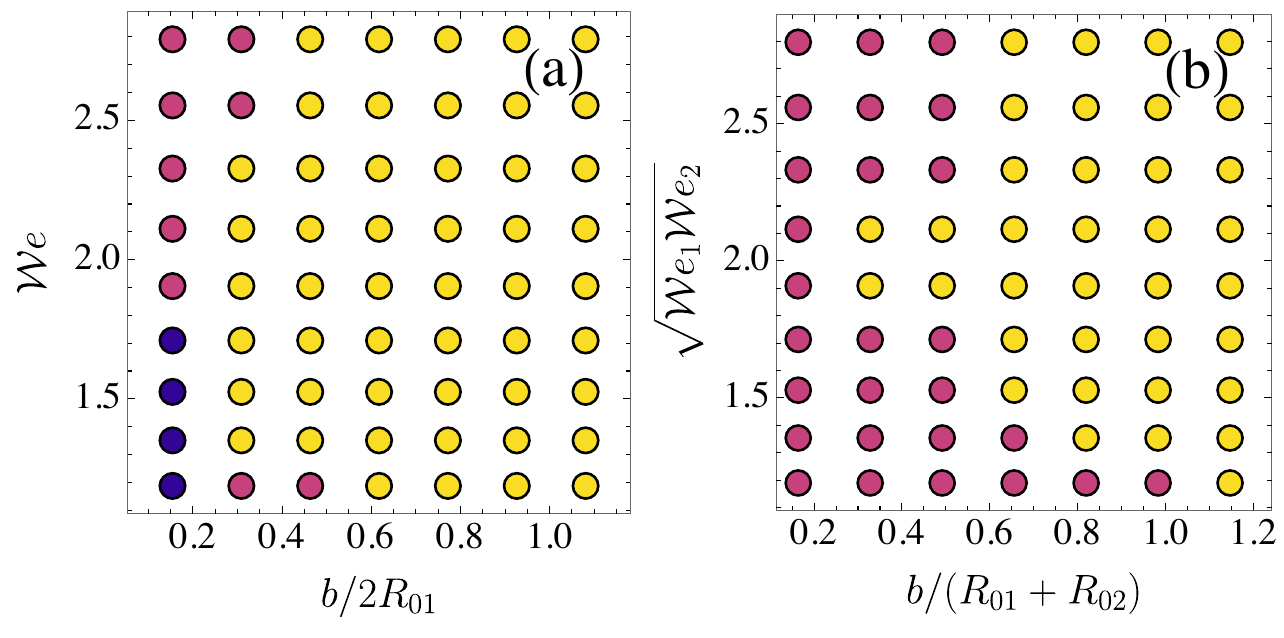}
    \caption{ {\bf Phase diagrams yielding different configurations  of topological defects inside the cloud  as a result of a binary collision.} Twin vortex rings [Fig.~\ref{fig:Fig2}(a)]  (dark blue, \tikzcircle[fill=blue!70!black]{3pt}), one vortex with a twisted ring  [Fig.~\ref{fig:Fig2}(b) (magenta, \tikzcircle[fill=magenta!80!black]{3pt}) and twin vortices  [Fig.~\ref{fig:Fig2}(c)] (yellow, \tikzcircle[fill=yellow]{3pt}).An imbalance in droplet size stabilizes the formation of a single vortex with a twisted ring at low energy and small impact parameter. Increasing either of these parameters results in the formation of two vortices. (a) Initial colliding droplets are the same size $N_1=N_2=1.81\times 10^5$, $R_{01}=3.18 \xi$. (b) Initial colliding droplets have different sizes, with a ratio $N_1/N_2=0.73$, $N_2=1.81\times 10^5$, $R_{01}=2.81 \xi$, $R_{02}=3.18\xi$. The window of time evolution is  $\tau_{\mathrm{evo}}/\tau\in [0,20.1 ] $ for (a) and (b). No topological singularities are observed for $\mathcal{W}e<1$.
        }
    \label{fig:Fig3}
\end{figure}

    \emph{Setup proposal.}   Experimental setups for binary quantum droplet collisions have been implemented in the past via double-well potentials~\cite{Ferioli2019}. A long enough  lifetime of such systems can be guaranteed by their adequate heteronuclear composition~\cite{Derrico2019}. The droplets collide after they are imprinted with a momentum kick, here described by plane-wave factors in opposite directions with wave vector $\mathbf{k}_0 = k_0\hat{e}_0$. They are separated by a vector $\mathbf{d} = 2d_0 \hat{e}_0 + b \hat{e}_\perp$ \footnote{In the reported data typical values of these quantities were {   
    $d_0 = 3.79 \mu m = 3.30 \xi$;
    $b \in [0.98 \mu m, 8.86 \mu m] = [0.86 \xi, 7.78 \xi]$; $k_0 \in [0.30 \mu m^{-1}, 0.48 \mu m] = [0.36 \xi^{-1}, 0.42 \xi^{-1}]$ }}
  i.e., generating a non-frontal collision for $b\neq0$, see Fig.\ref{fig:Fig1}(a) \respch{(see the Supplemental Material~\cite{SM})}.
  
  The dynamics of the droplets are governed by the extended Gross-Pitaevskii equation~\cite{Petrov2015, MALOMED2019108, guo2021, VanLoock2026, Alba2021, Photonics},
\begin{equation}\label{eq1}
\begin{aligned}
i \hbar \frac{\partial \psi_\alpha}{\partial t} = \Big(
-  \frac{\hbar^2}{2 m_\alpha} \nabla^2  & + g_{\alpha \alpha}  \lvert \psi_\alpha \rvert^2 + g_{\alpha \beta}  \lvert \psi_\beta\rvert^2    \\
&
+(
                g_{\alpha \alpha}^{\mathrm{Q}}\lvert \psi_\alpha \rvert^2  
             + g_{\alpha \beta}^ {\mathrm{Q}}\lvert \psi_\beta \rvert^2 
)^{3/2}
 \Big) \psi_\alpha,
\end{aligned}
\end{equation}
where for each BEC component $\alpha$ the order parameters is $\psi_\alpha$, the mass of each atom is $m_\alpha$, $g_{\alpha \alpha}$, $g_{\alpha \beta}$  are the coupling strengths arising from the low energy s-wave interactions and $g_{\alpha \alpha}^{\mathrm{Q}}$, $g_{\alpha \beta}^{\mathrm{Q}}$ incorporate Lee-Huang-Yang corrections due to quantum fluctuations, see~\footnote{
  $g_{\alpha\chi}^{\mathrm{Q}}= \left( \frac{4 }{3 \pi^2} \frac{g_{\alpha \alpha} m^{3/5}_\alpha}{\hbar^3} \right)^{2/3} m^{3/5}_\chi g_{\chi \chi}$ 
  }. 
Note that, in general, $\psi_a \neq \psi_b$. 

 \respch{\emph{Natural units.} For droplets that satisfy the condition $N_a /N_b = \sqrt{g_{bb}/g_{aa}} $ with $N_{\alpha}$ the number of atoms of the $\alpha=a,b$ species in the mixture, the natural unit of length is given by \cite{Petrov2015, Semeghini2018},
\begin{equation}\label{eq:long_natural_petrov}
 \xi_{a} = \hbar\sqrt{\frac{3}{2}\frac{\sqrt{,g_{bb}}/m_a + \sqrt{g_{aa}}/m_b}{\vert \delta g\vert \sqrt{g_{aa}}n_a^{(0)}}},\; \delta g = g_{ab} + \sqrt{g_{aa}g_{bb}}.\nonumber
\end{equation}
Where the equilibrium density is~\cite{Petrov2015},
\begin{equation}
n_a^{(0)} = \frac{25\pi}{1024} \frac{1}{(1+ (m_b/m_a)^{3/5} \sqrt{g_{bb}/g_{aa}})^5}\frac{1}{\mathrm{a}_{\alpha\alpha}^3}\frac{\delta g^2}{g_{aa}g_{bb}}.\nonumber
\end{equation}
The natural unit of time is
\begin{equation}
\tau_a =\frac{3\hbar}{2}
\frac{\sqrt{g_{aa}} + \sqrt{g_{bb}}}{\vert g_{ab} + \sqrt{g_{aa}g_{bb}}\vert \sqrt{g_{aa}}n_a^{(0)}}.\nonumber
\end{equation}
}

\respch{In our case, mixtures of $^{41}$K and $^{87}$Rb atoms are considered with realistic experimental parameters~\cite{Derrico2019}.} Scattering lengths in the reported results are: $\mathrm{a}_{\Potasio \Potasio} =62.0 a_0$, $\mathrm{a}_{\Rubidio \Rubidio} =100.4 a_0$, $\mathrm{a}_{\Potasio \Rubidio} =-82.0 a_0$, where $a_0$ is the Bohr radius and $g_{\alpha\beta}= 2\pi\hbar^2\mathrm{a}_{\alpha\beta}(1/m_\alpha +1/m_\beta)$. The  
natural units of the system are taken as: length $\xi=\xi_K=1.14 \mu$m and time $\tau=\tau_K= 1.184$ms.
The number of atoms of each atomic species is chosen so that the resulting droplets are in the incompressible regime \cite{Alba2021}.

Classical droplet dynamics identifies the Weber number $\mathcal{W}e= \mathcal{K}/{R^2_0 \sigma}$ as a natural reference energy scale for exploring the outcomes of droplet collisions \cite{FrohnBook2000}. This number relates the translational kinetic energy of the droplet  $\mathcal{K}$ to an estimate of its surface energy, calculated as the product of its squared radius $R_0$ and its surface tension $\sigma$ \cite{Alba2021}. Since $\mathcal{W}e$ depends on the droplet's characteristics, in a binary collision of non-identical drops, the geometric mean $\sqrt{\mathcal{W}e_1\mathcal{W}e_2}$ is the relevant parameter. By modifying the initial conditions, specifically $\mathcal{K}$ and the initial angular momentum $\vec{L} = \vec{L}_1 + \vec{L}_2$, using the impact parameter $b$, a competition is observed between the rotation of the resulting fluid as a whole, its deformation, and the eventual formation of vortices. For small Weber numbers, the droplets coalesce as already reported for head-on collisions~\cite{Alba2021, PhysRevResearch.3.043139, Ferioli2019} .
As $\mathcal{W}e$ increases, the final result for extended times, $t\gg\tau$, is two separate droplets. However, they remain together for $t\sim5 \tau$, and topological singularities can form within the cloud. 
This is faster than the expected lifetime for incompressible three-body droplet losses for the parameters of the heteronuclear mixture here reported, which is around $30-35 \tau$ ($\simeq 36-42$ ms) \cite{Alba2021}.
Additionally, for intermediate Weber numbers, $\mathcal{W}e\sim 1$, no vorticity is observed after the collision. Then, the description for the rotating superfluid can be done in terms of a time-dependent moment of inertia of the
system-- including the evolution of the principal axes-- and the conservation of angular momentum \cite{Photonics}.

\begin{figure*}[t!]
    \includegraphics[width=0.95\linewidth]{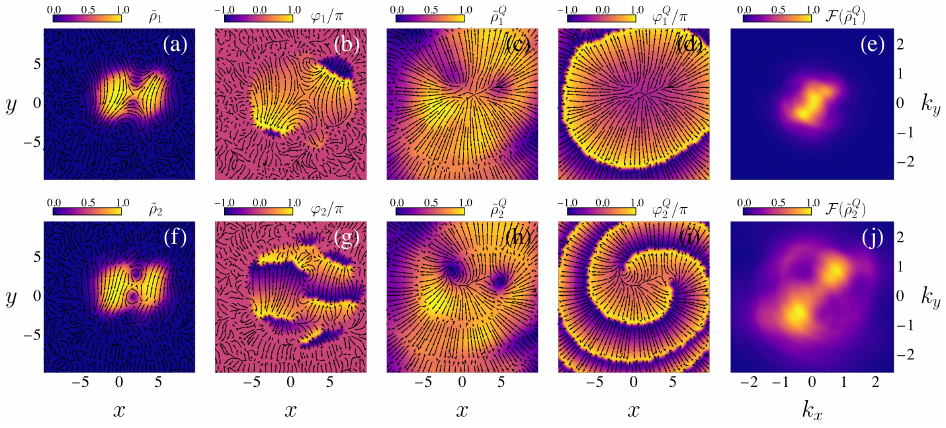}
    \caption{
    {\bf{Velocity fields, density and quantum phase distributions of the atomic cloud components for a collision that produces twin vortices.}} The heavy atomic component spins fast near a generated vortex filament, yielding minimal density regions in the vortex core. Density scaled to its maximum value (a,f) and phase (b,g) distributions for the light/heavy atomic components of the cloud with the  corresponding velocity fields (black arrows) after the collision. Density scaled to its maximum value (c,h) and phase (d,i) distributions for the light/heavy atomic components of the cloud with the corresponding velocity fields (black arrows) when the quench protocol is applied after the collision. Panels (e,j): Integrated Fourier transform preceded by a time of flight signal, of panels (c,h) for light/heavy atomic components. Parameters are the same as in Figs.~\ref{fig:Fig1} and \ref{fig:Fig2}~(c). The plane of viewing is the plane of collision defined by $\hat{e}_\parallel$. The units of $x,y$ are $\xi$, and the units of $k_x,k_y$ are $1/\xi$, $\tilde\rho_1,\tilde\rho_2$ in $8.03\times10^{-3}\xi^{-3}$, and $\tilde\rho_1^Q,\tilde\rho_2^Q$ in $3\times10^{-4}\xi^{-3}$, which are the maximum densities, with
    $\xi=1.14\mu$m.} 
    
  \label{fig:Fig4}
\end{figure*}

\emph{Topological defects.} An extensive computational search identified three types of topological defects that could be generated within the atomic cloud. Numerical methods~\cite{Strang1968, Bao2003, NUMERICALRECIPIES1992} are briefly described in the supplemental material~\cite{SM}.  These are formed during the superposition of the incompressible droplets in non-frontal binary collisions with Weber number $\mathcal{W}e \gtrsim 1.2$. The topological singularities are clearly observed in the heavier atomic species.
For low energy and small impact parameter, there are twin vortex rings,  provided that the difference of the initial droplet sizes is negligible, Fig.~\ref{fig:Fig2}(a). The rings are precursors to vortices that traverse the cloud at higher energies and larger impact parameters. However, if the energy and size difference is moderate, a single vortex with a twisted ring typically emerges, as shown in Figs.~ \ref{fig:Fig1}  and \ref{fig:Fig2}(b). Intuitively, this can be understood as there is not enough energy or momentum available to break both rings. Increasing the angular momentum and the kinetic energy finally reaches the point at which twin vortices are the most stable outcome, as seen in Figs.~\ref{fig:Fig1}  and \ref{fig:Fig2}(c). The distribution of the topological defects found in terms of the Weber numbers and the impact parameter is shown in Fig.~\ref{fig:Fig3} for collisions involving identical (a) and non-identical (b) droplets. Qualitatively similar results are found  with other similar conditions,
provided that the droplets are incompressible enough \respch{
(see the Supplemental Material~\cite{SM})}.

\emph{Key mechanisms.} 
The phenomenology of the dynamical generation of a wide variety of topological excitations during binary collisions can be traced to the quantum hydrodynamic properties of the gas components. Writing the kinetic quantum energy term of the Hamiltonian as
\begin{equation}\label{eq:MadelungTrans}
\frac{-\hbar^2 }{2 m_\alpha} \nabla^2 \psi_\alpha = \frac{m_\alpha \lvert \mathbf{v}_\alpha \rvert^2}{2}  -\frac{\hbar^2 }{2 m_\alpha}\frac{\nabla^2 \sqrt{\rho_\alpha}}{\sqrt{\rho_\alpha}}, \quad  \mathbf{v}_\alpha = \frac{\hbar}{m_\alpha} \nabla \varphi_\alpha,
\end{equation}
with a local velocity field $\mathbf{v}_\alpha$, one identifies the first term as the usual kinetic energy contribution, and the second as the quantum pressure for the $\alpha$-component. 
Both terms in Eq.~(\ref{eq:MadelungTrans}) exhibit a mass dependence that varies inversely with atomic mass; this renders vortex nucleation energetically more favorable in Rb and, in the case of K, promotes the formation of a non-circulating density decrease in the vicinity of a Rb vortex. This smooths out large density variations that would otherwise result in significant quantum-pressure contributions to the energy~\cite{ParkerLibro, VorticesAncilotto, barranco2023}. Vortex nucleation occurs at the periphery of the resulting droplet, where density is lower, and strong variations are energetically more favorable than in the denser center. 

  Unlike other BEC systems, in heteronuclear quantum droplets, the resulting coreless vortex yields a null contribution to the interspecies density-density interaction near it. All these energetic features are directly accounted for in our computational approach. The time-dependent order parameter of each gas component, Rb or K,  encodes a local redistribution of the energy, linear and angular momenta during the binary scattering process. Nevertheless, within numerical precision, these dynamical variables are conserved during the unitary evolution. 

\emph{Details of the expected experimental outcomes}. An illustrative example of the behavior of the densities, velocity field, and phase during a non-frontal binary collision is shown in Figs.~\ref{fig:Fig4}(a,b,f,g), which results in twin-vortex formation. The same parameters as in Figs.~\ref{fig:Fig1} and \ref{fig:Fig2}(c) are considered. The heavy component (Rb), shown in Figs.~\ref{fig:Fig4}~(f,g), exhibits an increasing velocity field close to the vortex filament. This can be understood as a consequence of the inverse dependence on the distance to the dislocation line of the gradient in the expression of the quantum hydrodynamic velocity $\mathbf{v}_{Rb}$. In contrast, the light component (K), Fig.~\ref{fig:Fig4}(a,b),  shows no phase dislocations in the fused cloud, i. e., no vorticity. 

The microscopic size of the topological excitations might make direct measurement challenging. In order to enhance the detection signals, a rapid quench~\cite{PhysRevLett.120.253002} of the interspecies scattering length via the associated magnetic fields  can be applied just after the droplet collision takes place ($t_{evo}\sim 10 \tau- 15 \tau$ in the illustrated examples, Fig.~\ref{fig:Fig4}(c,d,e,h,i,j)). By diminishing attractive interactions among the atoms, the gas is removed from the droplet and allowed to expand into the unbounded gas regime. If properly implemented, the size of the resulting mixture increases while the topological singularities are preserved. \respch{(see the Supplemental Material~\cite{SM})}. We show the results of the quench protocol for the spatial distribution of $\rho^Q_{\alpha}$, $\varphi^Q_\alpha$, with the corresponding velocity fields in Figs. \ref{fig:Fig4}(c,d) for the light component and (h, i)  for the heavy component. 
These simulations illustrate a substantial increase in the size of the  cores of topological singularities, making them suitable for direct imaging. The vortex structure endures the expansion, with the density depression and phase information preserved. A distinctive feature of a vortex is that the decrease in density near a phase singularity line also occurs in momentum space. Therefore, the system can be studied using a time-of-flight (TOF) protocol supplemented by a signal in the transverse momentum density function,
\begin{equation}
  \mathcal{F}(\rho_{\alpha}) (k_x,k_y)  = \int |\mathcal{F}(\psi_{\alpha})|^2 d k_z.
  \label{eq3}
\end{equation}
In Figs.~\ref{fig:Fig4}(e) and (j), the predicted  Fourier space $\mathcal{F}-$densities for the two atomic species corresponding to the collision described in Fig.~2c are illustrated. The resulting $\mathcal{F}(\tilde{\rho}^Q_1)$ for the light atoms shows no local depressions but rather a peak centered at the origin. This peak arises from the initial conditions, which tend to maximize $\mathcal{F}$ at the origin, with approximate $(k_x,k_y)\leftrightarrow (-k_x,-k_y)$ symmetry. For the heavier atoms, the presence of the vortices creates a competition between this tendency and the density depletion near the vortices. The latter possess twisted singularity lines. The result is the appearance of a pair of quasi-symmetric peaks in the $^{87}$Rb  $\mathcal{F}(\tilde{\rho}^Q_2)$ density, alongside three depressions surrounded, in this case, by a ring-like structure. These features serve as distinctive signatures of the absence (lighter atoms) or presence (heavier atoms) of topological defects. It should be noted that this procedure does not allow for the direct resolution of the topological defect's three-dimensional structure. To achieve 3D resolution, similar TOF procedures employing multiple imaging beams with different orientations could prove useful \cite{dallibard2002}.
Note that for the collision parameters under consideration, vortex nucleation occurs at early stages ($t_{evo} \sim$ 12–18 ms), which is when the imaging procedure would begin.
 
In this Letter,  we identify the scenarios and mechanisms underlying the dynamical generation of topological defects in non-frontal binary collisions of quantum droplets. 
Specific structural and kinematic conditions are required to nucleate topological singularities seeded by emergent density and phase deformations during the time interval when the quantum droplets superpose. Interestingly, at later times after the collision, the atomic cloud reconfigures into two separate quantum droplets. The mechanisms identified here for achieving nontrivial dynamical outcomes are general and could be further explored in other nuclear and atomic systems in the superfluid regime.  The theoretical findings suggest a broad, feasible parameter region in which fundamental aspects of the formation and evolution of the reported excitations can be observed in experiments involving heteronuclear Bose-Bose mixtures  of ultracold atomic gases. The emergence of cloud rotation, capillary waves, phase dislocations, and  vortices can be controlled by the intrinsic characteristics of the droplets and the initial conditions.
It has also been shown that clear experimental signatures of these behaviors are expected via time-dependent protocols where information in configuration and momentum spaces can be efficiently retrieved.
Higher Weber numbers, leading to additional fragmentation of the cloud, might give rise to other topological defects and, possibly, new dynamical regimes to be explored in the future.

\begin{acknowledgments}
We thank J. A. Seman and F. J. Poveda-Cuevas for useful discussions. Computational simulations were performed in the HPC server LSCSC-LANMAC. This work is partially supported by the grants UNAM-DGAPA-PAPIIT: IN118823, UNAM-DGPA-PAPIIT: IG101826, UNAM-CIC: CONAHCYT/SECIHITI: LNC-2023-51,  the (Polish) National Science Center Grant No. 2022/45/B/ST2/00358, as well  grants from NVIDIA and
usage NVIDIA RTX 6000 Ada.
\end{acknowledgments}

\bibliography{main_ACCEPTED.bbl}

\clearpage

\onecolumngrid
\begin{center}
\textbf{\large Supplemental material for: Collision of Quantum Worlds: Vorticity Induced by Non-frontal Collisions of Quantum Droplets}
\\
\bigskip
{Jose E. Alba-Arroyo$^{1,2}$, Santiago F. Caballero-Benitez$^{3}$ and Rocio J\'auregui$^{2}$ }
\\
\bigskip
{\it $^1$Faculty of Physics, Warsaw University of Technology, Ulica Koszykowa 75, 00-662 Warsaw, Poland}
\\
{%
{\it $^{2}$Departamento de F\'{\i}sica Cu\'antica y Fot\'onica, Instituto de Física, Universidad Nacional Autónoma de México, Ciudad de México 04510, Mexico}
}%
 \\
{%
\it $^{3}$Departamento de F\'{\i}sica Cu\'antica y Fot\'onica, LSCSC-LANMAC, Instituto de Física, Universidad Nacional Autónoma de México, Ciudad de México 04510, Mexico
}%
\end{center}

 Here we present model details used in the Extended Gross-Pitaevskii frame, specific analytical expressions for the initial conditions used in our simulations, the specific form of the sweeping protocol implemented in the calculations and suggested for an experimental realization, and a brief description of the computational simulations methodology. 
\\
\\

\twocolumngrid

\subsection{Model Details}
Consider a binary mixture of Bose gases composed of $N^{(a)}$ and $N^{(b)}$ atoms of each species. Let us assume that the  state of the system is approximately described by the Hartree many-body wavefunction 
\begin{equation}
\begin{aligned}
 & \Psi_N(\vec r_1,...,\vec r_{N^{(a)}};\vec r^\prime_1,...,\vec r^\prime_{N^{(b)}};t) 
= \\
&\Big[\prod_{i=1}^{N^{(a)}}\chi^{(a)}(\vec r_i;t)\Big]\Big[\prod_{i=1}^{N^{(b)}}\chi^{(b)}(\vec r^\prime_i;t)\Big].
\end{aligned}
\end{equation}
As order parameters, we consider
\begin{equation}
\psi^{(\alpha)}(\vec r,t) = \sqrt{N^{(\alpha)}} \chi^{(\alpha)}(\vec r,t), \quad\quad  \alpha = a, b,\label{eq:normalizacion}
\end{equation}
With $\chi_\alpha$ normalized to one, and $\psi^{(\alpha)}$ evolves according to the equations
\begin{subequations}\label{eq:evolU}
\begin{align}
&i\hbar \partial_t \psi^{(\alpha)}(\vec r,t) 
= \hat H_0^\alpha \psi^{(\alpha)}(\vec r,t) + \hat{ U}^\alpha \psi^{(\alpha)}(\vec r,t)   \label{eq:evol}\\
&\hat H_0^\alpha = -\frac{\hbar^2}{2m_\alpha}\nabla^2 + V^\alpha_{ext}(\vec r)\,\, , \label{eq:H0}\\
&\hat{ U}^\alpha \psi^{(\alpha)}(\vec r,t)
= \int d^3 r^{\prime}\,
   \Bigg[
   \mathcal{U}_{\alpha\alpha}(\vec r, \vec r^\prime;\rho^{(\alpha)}(\vec r^\prime,t))  \notag\\
& + \mathcal{V}_{\alpha\beta}(\vec r, \vec r^\prime;\rho^{(\alpha)}(\vec r^\prime,t),\rho^{(\beta)}(\vec r^\prime,t))
   \Bigg]\psi^{(\alpha)}(\vec r,t) \label{eq:U}
\end{align}
\end{subequations}

In Eq.~(\ref{eq:evol}), $V^\alpha_{ext}(\vec r)$  is the external potential and $\rho^{(\alpha)}=\psi^{(\alpha)*}\psi^{(\alpha)}$ the density of  $\alpha$--atoms. 
 The real functionals $\mathcal{U}_{\alpha\alpha}$  and $\mathcal{V}_{\alpha\beta}$ in the integral operator $\hat{U}^\alpha$ may incorporate, in an effective way, interactions beyond the standard mean field approximation. The functional $\mathcal{V}_{\alpha\beta}$  represents the interaction between different species,
 i.e. $\alpha\neq\beta$. In the case of EPGE, the LHY term is included, so that the density-dependent interactions are superpositions of contact terms~\cite{Petrov2015},
\begin{subequations}
\begin{align}
\mathcal{U}_{\alpha\alpha} &= g_{\alpha\alpha}\rho^{(\alpha)}(\vec r^\prime) \delta^3(\vec r - \vec r^\prime)\\
\mathcal{V}_{\alpha\beta}&= \Big[g_{\alpha\beta}\rho^{(\beta)}(\vec r^\prime)  +\frac{4m_\alpha^{3/5}g_{\alpha\alpha}}{3\pi^2\hbar^3} \\
& \Big(m_\alpha^{3/5}g_{\alpha\alpha}\rho^{(\alpha)}(\vec r^\prime)+m_\alpha^{3/5}g_{\beta\beta}\rho^{(\beta)}(\vec r^\prime)\Big)^{3/2}\Big]\delta^3(\vec r - \vec r^\prime),
\end{align}
\end{subequations}
with $m_\alpha$ the atomic mass and the coupling constants $g_{\alpha\beta}= 2\pi\hbar^2\mathrm{a}_{\alpha\beta}(1/m_a +1/m_b)$. Our main interest is to use values as in experimental realizations such that $g_{aa}> 0$, $g_{bb} > 0$, and $g_{ab}<0$ $(\delta g  = \sqrt{ g_{aa} g_{bb}} - g_{ab} \lesssim 0)$ \cite{Semeghini2018}.

The equation of motion with the previous assumptions is 
\begin{equation}\label{eq:TDEGPEE}
\begin{aligned}
&i \hbar \frac{\partial \psi_\alpha}{\partial t} = \left(
-  \frac{\hbar^2}{2 m_\alpha} \nabla^2  + g_{\alpha \alpha}  \lvert \psi_\alpha \rvert^2 + g_{\alpha \beta}  \lvert \psi_\beta\rvert^2   +  \right. \\
&\left.
\frac{4 }{3 \pi^2} \frac{g_{\alpha \alpha} m^{3/5}_\alpha}{\hbar^3}
(
                m^{3/5}_\alpha g_{\alpha \alpha} \lvert \psi_\alpha \rvert^2  
             +  m^{3/5}_\beta g_{\beta \beta} \lvert \psi_\beta \rvert^2 
)^{3/2}
 \right) \psi_\alpha,
\end{aligned}
\end{equation}
which coincides with Eq.~(\ref{eq:TDEGPEE}) from the main text.

The stationary ground state solutions of Eq.~(\ref{eq:evolU}) define the chemical potentials $\mu_\alpha$,
\begin{equation}\label{eq:STATIC}
\psi_\alpha (\vec r,t)= e^{-i\mu_\alpha t/\hbar}\phi^{(\alpha)}_0(\vec r),\quad\quad  \alpha = a, b.
\end{equation}

The stationary solutions for this equation have been widely discussed in the literature \cite{MALOMED2019108}. In particular, in a previous article \cite{Alba2021} we calculated a full parametrization of heteronuclear and homonuclear quantum droplets for $^{39}$K-$^{39}$K, and $^{41}$K-$^{87}$Rb systems. This description allowed us to calculate the radius $R_0$ of the droplet, its saturation density, and its density profile  as a function of the number of atoms. We also explored the dynamics of self-evaporation and considered the dynamics of frontal collisions and a scheme to parametrize their outcome as a function of dynamical quantities such as surface tension ($\sigma$).

\subsection{Initial conditions}\label{ssec:InitialCond}

Consider the collision of two generally different mixed-quantum droplets ($i=1,2$). Each of these with an atomic population $N_{\alpha i}$, with $\alpha=a,b$, denoting the atomic species. They both start with a momentum defined by a plane wave factor $\vec k_0 = k_0\hat{e}_0$, and are separated by a vector $\vec d = 2d_0 \hat{e}_0 + b \hat{e}_\perp$,
\\
\begin{eqnarray}\label{cond_inic_lab}
&\Psi&(\vec r, t=0) =\Psi_1(\vec r) + \Psi_2(\vec r)  \nonumber\\
&=&\begin{pmatrix}\psi_{a_1}(\vec r + \vec d_0)\\ \psi_{b_1}(\vec r+ \vec d_0)\end{pmatrix} e^{i\vec k_0\cdot \vec r/2}
\\
&+&\begin{pmatrix}\psi_{a_2}(\vec r - \vec d_0 + b \hat{e}_\perp)\\ \psi_{b_2}(\vec r- \vec d_0+ b \hat{e}_\perp)\end{pmatrix} e^{-i\vec k_0\cdot \vec r/2}.  \nonumber
\end{eqnarray}
Separation $d_0 \ggg R^{i=1,2}_0$ is chosen to numerically guarantee a negligible initial overlap of the droplets.

In the center-of-mass frame, (\ref{cond_inic_lab}) can be rewritten as
\begin{eqnarray}\label{cond_inic_cm}
&&\Psi(\vec r, t=0) = \begin{pmatrix}\psi_{a_1}(\vec r + \vec d_0)\\ \psi_{b_1}(\vec r+ \vec d_0)\end{pmatrix} 
e^{i F_1 \vec{\respch{k}}_0 \cdot \vec{r}}
+ \nonumber\\
&&\begin{pmatrix}\psi_{a_2}(\vec r - \vec d_0 + b \hat{e}_\perp)\\ \psi_{b_2}(\vec r- \vec d_0+ b \hat{e}_\perp)\end{pmatrix} 
e^{-i F_2 \vec{\respch{k}}_0 \cdot \vec{r}}, \label{eq:initial}
\end{eqnarray}
with $F_1 =  \sqrt{3} N_{b2} k_0  /(N_{b2}+N_{b1})$, $F_2 =  \sqrt{3} N_{b1} k_0 /(N_{b2}+N_{b1})$.

In this case, the initial linear and angular momenta for species $\alpha$ per number of particles $N_\alpha$ take the compact form:
\begin{equation}\label{eq:INITIAL_PL}
\begin{aligned}
&\vec{P} (t=0) = 0 \\
&\mathcal{K}^{tras} (t=0) = \frac{\hbar^2 k^2_0 }{4 }\sum_{i=1,2} \frac{1}{N_{bi}^2} \left( \frac{N_{ai}}{m_a} +  \frac{N_{bi}}{m_b}  \right) \\
&\vec{L}_\alpha (t=0)= -  \sqrt{3}  \hbar b_0 k_0  \frac{N_{b1}  }{N_{b2} + N_{b1}}  N_{\alpha 2} \hat{e}_0 \\
&\vec{L} (t=0)=  \vec{L}_a (t=0) + \vec{L}_b (t=0)= \\
& - \hbar  \sqrt{3} N_{b1} b_0 k_0  \frac{N_{b2} + N_{a2}}{N_{b2} + N_{b1}} \hat{e}_0
\end{aligned}
\end{equation}
For comparison, the largest amount of angular momentum that a large droplet consisting of the populations of the two colliding droplets could get if a vortex formed through its diameter is $L_\alpha (t=0) = \hbar \left( N_{\alpha1} + N_{\alpha 2} \right) $

\subsection{Computational simulations methodology.}
The general procedure started with the calculation of the initial ground-state wave functions for each droplet, including their finite lifetime due to three-body losses in homonuclear and heteronuclear atomic configurations.
This allowed the identification of feasible set-ups where the collision could be implemented and characterized.
Parameter values that define the binary collision between the heteronuclear droplets with Weber numbers
$We$ for identical droplets and $\sqrt{We_1We_2}$ for non identical droplets in the range [1., 3.0], 
aditionally $b/(2R_{01})$ or $b/(R_{01}+R_{02})\in$ [0.1, 1.2] and the time intervals 
$t_{evo}/\tau \in$ [0, 20.1] were explored as illustrated in Fig.3.
The $We$ parameter value was introduced via the adequate phase factor in the order parameter of each droplet. The  unitary evolution of the system  in 3D was numerically implemented using pseudospectral methods.  Systematic checks regarding the behavior of norm and energy, as
well as the stability of the results under  diverse spatial regions using periodic boundary conditions, were performed across the entire reported parameter space in order to avoid numerical artifacts and optimize code execution time. For each parameter choice and at various time steps,  the density
distribution of the sum of the components and of each individual component (Rb/K) of the
droplets was analyzed via 3D imaging, considering resolutions ranging from 64$^3$
to 256$^3$ points in our simulation boxes. In addition to these qualitative 3D explorations, we analyzed
the behavior of the order parameter in several 2D planes defined by different axes;
in particular  density components  and wave-function phases were evaluated. From these
quantities,  projections of the system’s current vector field and densities of the
angular and linear momentum components were evaluated, besides the topological charges associated with the observed topological defects. These exhaustive systematic explorations—both qualitative
and quantitative—clearly demonstrated that the observed vortex dynamics represent regular
behavior that is insensitive to small deviations in the computational simulations. 
 Stability with respect to simulation resolution was also optimized to reduce the execution time of our codes. Finally, some simulations of the collision dynamics accounting for three-body losses were conducted: within the simulation time windows considered, the behavior of the resulting topological excitation remained robust. Finally, selected measurement signals were identified and illustrated in Fig.4(e) and (j). In addition to the atomic proportions reported in this work, other configurations exhibiting similar general behavior were considered and will be discussed elsewhere.

\subsection{Specific form for the quenching protocol.}
The interspecies scattering length is taken to be time dependent according to~\cite{PhysRevLett.120.253002}:
\begin{equation}
\begin{aligned}
& a_{12} (t_r) =  a_{12}(t_0) + \sigma(t_r,\Delta t)(a_{12}(t_0)-a_{12}(t_1)),  \\
& \sigma(t_r,\Delta t) = 0.5 \left(1 + \tanh( 3 \tan(\frac{\pi t_r}{ \Delta t}-\frac{\pi}{2})) \right),
\end{aligned}
\end{equation}
 changing $a_{12}(t_0) = -82 a_0$ to $a_{12}(t_{1}) = -73.6 a_0$ in a span of $\Delta t = t_{1} - t_0 = 2.7 \tau$, larger than the system's natural unit of time but much smaller than the timescale of the collision kinematics. This change in the scattering length corresponds to a ramp of a Feshbach magnetic field $\Delta \text{B} \simeq 218.96~\text{G/s}$.

\subsection{Numerical methods}
We numerically solve Eq.~(\ref{eq:TDEGPEE}) with the initial condition Eq.~(\ref{cond_inic_cm}) by applying the time-evolution operator
\begin{eqnarray}
\Psi(\vec r, t + \Delta t) = \exp(-i \mathcal{H} \Delta t / \hbar ) \Psi(\vec r, t + \Delta t),
\end{eqnarray}
where $\mathcal{H}$ is indentified with the right side of Eq.~(\ref{eq:TDEGPEE}). We implemented a Strang decomposition (Split-Step Method), which is accurate to second order in the time step $\Delta t$ 
order in the time step \cite{Strang1968}-\cite{Bao2003}:
\begin{eqnarray}
 &\exp(-i \mathcal{H} \Delta t / \hbar ) \approx 
 \exp(-i \Delta t \mathcal{V}  / (2\hbar) ) \times\\ \nonumber
&\exp(-i \Delta t \mathcal{\hat{K}}  / \hbar ) 
 \exp(-i \Delta t \mathcal{V}  / (2\hbar) ).
\end{eqnarray}
Where $\mathcal{V}$ is the Hamiltonian that acts in configuration space accounting for external potential and interactions, and $\mathcal{\hat{K}} $ is  the kinetic energy operator, 
\begin{equation}
\mathcal{K} =-\frac{\hbar^2}{2}\sum_{i=1,2}\int d^3r \Psi_i^\dagger(\vec r)\begin{pmatrix}  \frac{\nabla^2}{m_{a_i}}  & 0\\ 0 & \frac{\nabla^2}{m_{b_i}}  \end{pmatrix} \Psi_i(\vec r),\label{eq:K}
\end{equation}
which is evaluated in $k$-space by means of Fast Fourier transforms (FFT). The FFT is implemented according to \cite{NUMERICALRECIPIES1992}, and the entire algorithm has been developed in Fortran. Additional tests were performed in simulations with GPUs using Julia-based code.

\end{document}